\documentclass[epj]{webofc}
\usepackage[varg]{txfonts}   % Web of Conferences font
\woctitle{MESON2016 - the 14$^\textrm{th}$ International Workshop on Meson Production, Properties and Interaction}
\begin{document}
\selectlanguage{english}
\title{Measurement of hadron cross sections with the SND detector}

% insert email only for speaker/presenter
\author{V.~P.~Druzhinin\inst{1,2}\fnsep\thanks{\email{druzhinin@inp.nsk.su}} \and
        M.~N.~Achasov\inst{1,2} \and
        A.~Yu.~Barnyakov\inst{1,2} \and
	K.~I.~Beloborodov\inst{1,2} \and
	A.~V.~Berdyugin\inst{1,2} \and
	A.~G.~Bogdanchikov\inst{1} \and
	A.~A.~Botov\inst{1} \and
	T.~V.~Dimova\inst{1,2} \and
	V.~B.~Golubev\inst{1,2} \and
	L.~V.~Kardapoltsev\inst{1,2} \and
	A.~G.~Kharlamov\inst{1,2} \and
	A.~A.~Korol\inst{1,2} \and
	S.~V.~Koshuba\inst{1} \and
	D.~P.~Kovrizhin\inst{1,2} \and
	A.~S.~Kupich\inst{1} \and
	K.~A.~Martin\inst{1,2} \and
	A.~E.~Obrazovsky\inst{1} \and
	E.~V.~Pakhtusova\inst{1} \and
	S.~I.~Serednyakov\inst{1,2} \and
	D.~A.~Shtol\inst{1,2} \and
	Z.~K.~Silagadze\inst{1,2} \and
	I.~K.~Surin\inst{1,2} \and
	Yu.~V.~Usov\inst{1,2} \and
	A.~V.~Vasiljev\inst{1,2}
% comment out the next line if not needed
       \\The SND Collaboration
}

\institute{Budker Institute of Nuclear Physics, Novosibirsk, 630090, Russia
\and
          Novosibirsk State University, Novosibirsk, 630090, Russia
          }

\abstract{New results on exclusive hadron production in $e^+e^-$ annihilation obtained in experiments with the SND detector at the VEPP-2M and VEPP-2000 $e^+e^-$ colliders are presented.
}
\maketitle
\section{Detector and experiment}
The SND~\cite{det1,det2,det3,det4} is the general purpose nonmagnetic detector.
Its main part is a spherical three-layer electromagnetic calorimeter
containing 1640 NaI(Tl) crystals. Directions of charged particles are measured
by a tracking system based on a nine-layer drift chamber.
The particle identification is provided by $dE/dx$ measurements
in the tracking system and a system of aerogel Cherenkov counters.
Outside the calorimeter a muon detector is located.

SND collected data at the VEPP-2M~\cite{vepp2m} and VEPP2000~\cite{vepp2000}
$e^+e^-$ colliders. 
At VEPP-2M, data with an integrated luminosity of about 30 pb$^{-1}$ were 
recorded in 1996-2000 in the energy range 0.4--1.4 GeV. 
At VEPP2000, a wider energy interval, 0.3--2.0 GeV, is studied.
A 69 pb$^{-1}$ data sample was collected during 2010-2013. Currently the 
VEPP-2000 accelerator complex is under upgrade. Experiments with increased
luminosity are expected to be started in the end of 2016.

The main goal of the SND experiments is careful measurement of exclusive 
hadronic cross sections below 2 GeV, which are, in particular, needed
for the Standard Model calculation of the muon $(g-2)$ and running
$\alpha_{QED}$. Here we present new results on measurement of four exclusive
cross sections.
\section{Precise cross-section measurements}
The $e^+e^-\to \pi^0\gamma$ cross section is the third largest cross section
(after $e^+e^-\to \pi^+\pi^-$ and $\pi^+\pi^-\pi^0$) below 1 GeV.
From analysis of the $e^{+}e^{-}\rightarrow \pi ^{0}\gamma$ data in the
vector meson dominance (VMD) model, the widths of vector-meson radiative 
decays are extracted, which are widely used in 
phenomenological models.
The most accurate data on this process were obtained in experiments at 
the VEPP-2M $e^+e^-$ collider with the SND~\cite{snd1,snd2} and 
CMD-2~\cite{cmd} detectors. The SND results~\cite{snd1,snd2} are based on 
about 25\% of data collected at VEPP-2M. Here we present a new 
analysis~\cite{pi0g} using the full SND@VEPP-2M data sample.
\begin{figure}[ht]
\centering
\includegraphics[width=0.25\textwidth]{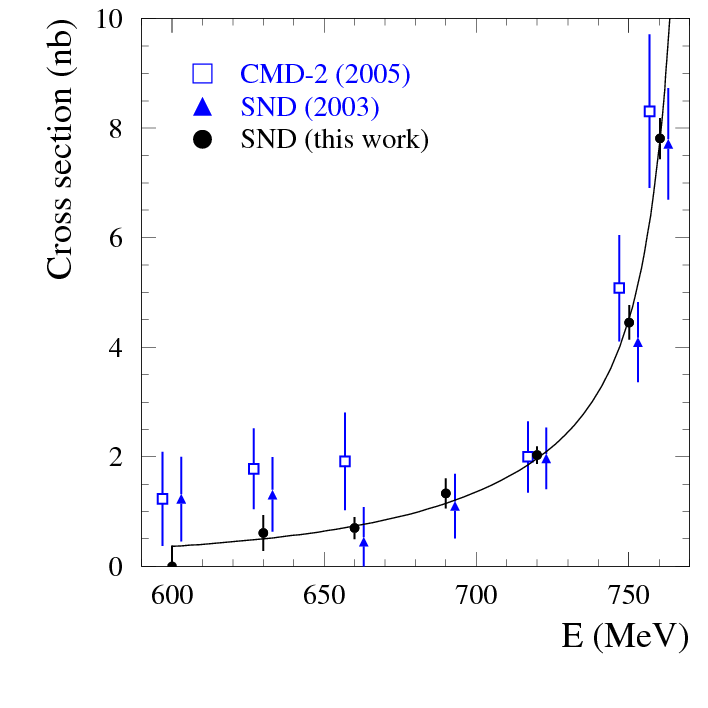}
\includegraphics[width=0.24\textwidth]{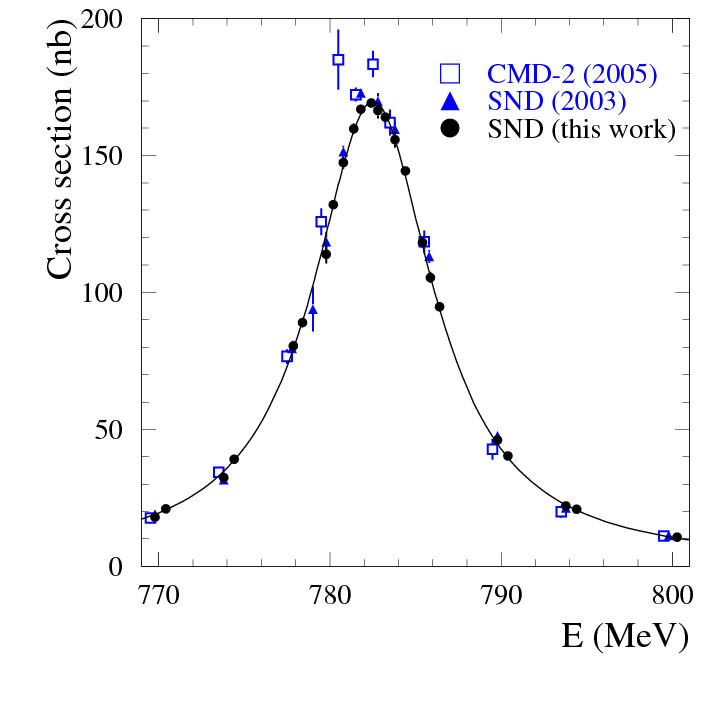}
\includegraphics[width=0.25\textwidth]{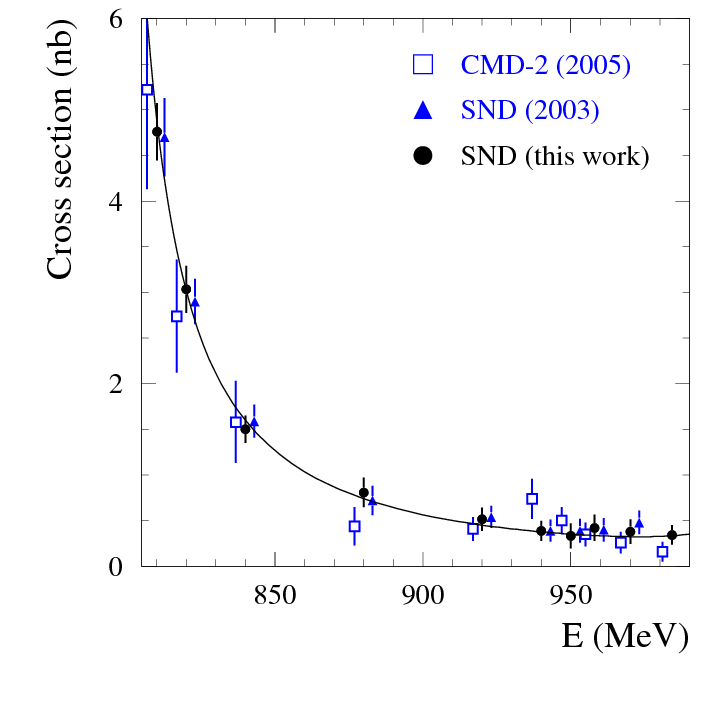}
\includegraphics[width=0.25\textwidth]{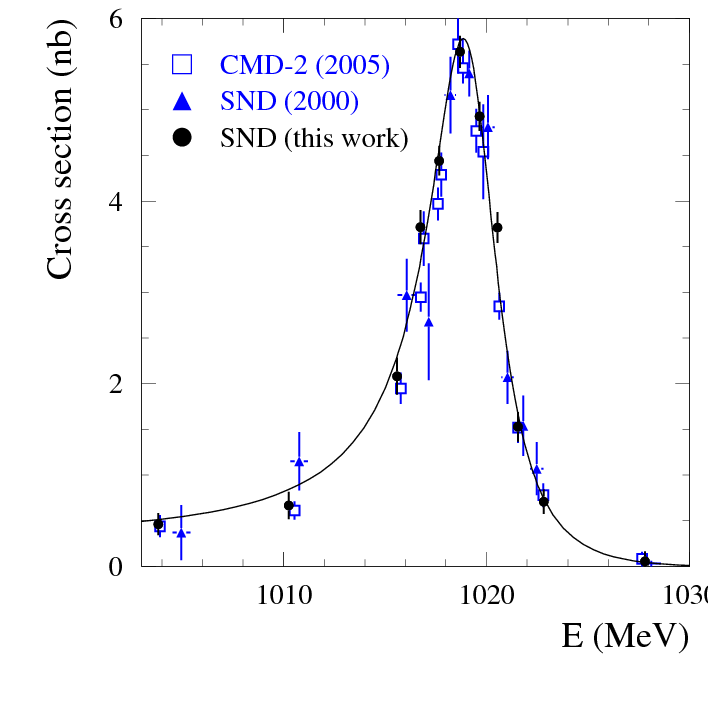}
\includegraphics[width=0.25\textwidth]{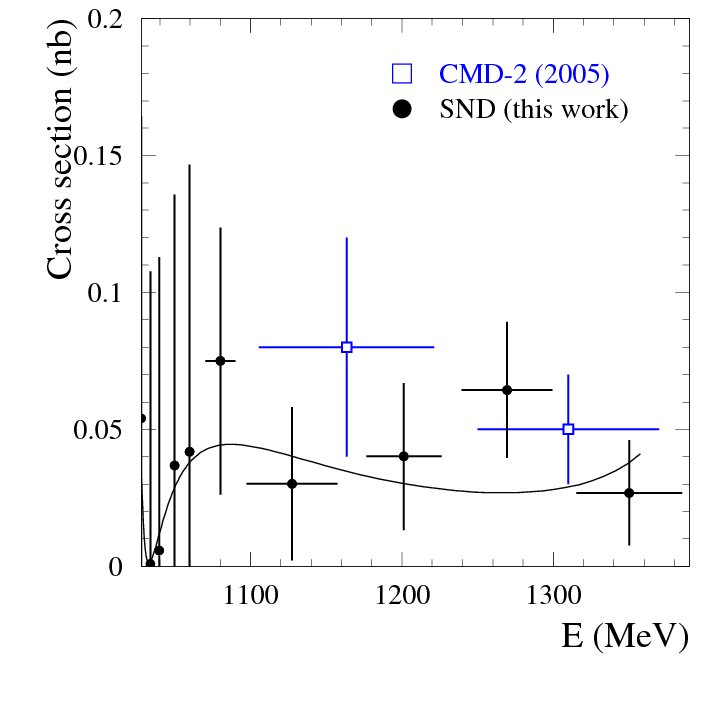}
\caption{The $e^+e^-\to \pi^0\gamma$ cross section measured by SND using
the full VEPP-2M data sample in comparison with the previous most accurate
measurements. The curve is the result of the VMD fit. Only 
statistical errors are shown. The systematic errors are 3.2\%, 3\%, and 6\% 
for SND (2000), SND (2003), and CMD-2 (2005) data, respectively. Our
systematic uncertainty at the $\omega$ and $\phi$ peaks is 1.4\%.
\label{fig5}}
\end{figure}

The measured $e^{+}e^{-}\rightarrow \pi ^{0}\gamma$ cross section shown in 
Fig.~\ref{fig5} agrees with previous SND~\cite{snd1,snd2} and CMD-2~\cite{cmd}
measurements within the systematic uncertainties, but is significantly 
more precise. 
From the fit to cross section data we obtain the products of branching 
fractions:
\begin{eqnarray}
B(\rho\to\pi^0\gamma)B(\rho\to e^+e^-)&=&
(1.98\pm0.22\pm0.10)\times10^{-8},\nonumber\\
B(\omega\to\pi^0\gamma)B(\omega\to e^+e^-)&=&
(6.336\pm0.056\pm0.089)\times10^{-6},\nonumber\\
B(\phi\to\pi^0\gamma)B(\phi\to e^+e^-)&=&
(4.04\pm0.09\pm0.19)\times10^{-7},
\end{eqnarray}
For the $\omega$ meson, we improve accuracy of the
product by a factor of 1.6. There is a tension between the value of the
ratio $B(\omega\to\pi^0\gamma)/B(\omega\to\pi^+\pi^-\pi^0)$ measured
by KLOE~\cite{kloe} and the same value obtained using VEPP-2M data. 
After our $\pi^0\gamma$ measurement this tension increases up to 3.4$\sigma$.
Our $\rho\to\pi^0\gamma$ measurement is most precise. The obtained branching
fraction $B(\rho\to\pi^0\gamma)=(4.20\pm0.52)\times10^{-4}$ is lower
than the PDG value by $1.8\sigma$, but agrees with the branching
fraction for the charged $\rho$ decay.
The result for the $\phi\to\pi^0\gamma$ has an accuracy comparable
with that of the PDG value. It is obtained in the fit, in which the relative 
phase between the $\phi$ and $\omega$ amplitudes ($\varphi_{\phi\omega}$) is 
taken from data on the $e^+e^-\to \pi^+\pi^-\pi^0$ process. In the fit with 
free $\varphi_{\phi\omega}$, 
$B(\phi\to\pi^0\gamma)$ is determined with about 20\% uncertainty.

\begin{figure}[ht]
\centering
\includegraphics[width=0.4\textwidth,trim=0mm 20mm 0mm 0mm]{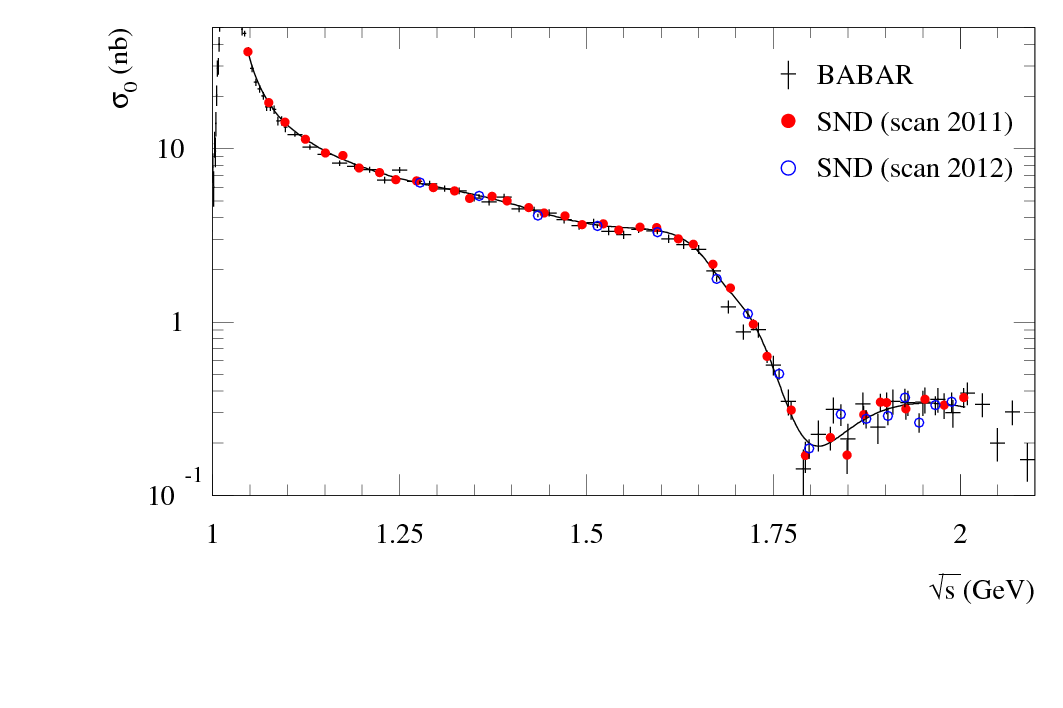}
\includegraphics[width=0.4\textwidth,trim=0mm 20mm 0mm 0mm]{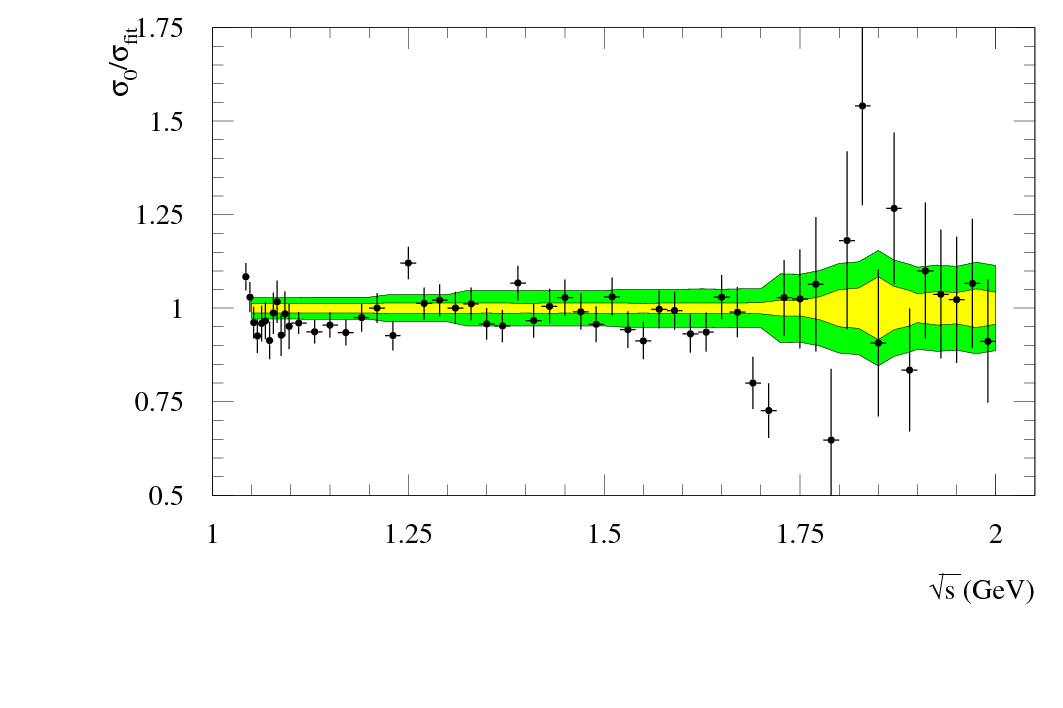}
\caption{Left panel: The $e^+e^-\to K^+K^-$ cross section measured 
by SND at VEPP2000 and in the BABAR experiment~\cite{babar}. The curve is 
the result of the VMD fit.
Right panel: The relative difference between the $e^+e^-\to K^+K^-$ cross 
sections measured by BABAR and the fit to the SND data.
The SND and BABAR systematic uncertainties are shown
by the light and dark shaded bands, respectively.
\label{fig8}}
\end{figure}

We also improve accuracy of $e^+e^-\to K^+K^-$ cross section measurement
in the energy range 1.05--2.00 GeV~\cite{KK}. Our result in comparison
with most precise previous measurement by BABAR~\cite{babar} is shown
in Fig.~\ref{fig8}
\section{Previously unmeasured cross sections}
Relatively precise inclusive data on $e^+e^-$ annihilation to hadrons exist 
above 2 GeV. Below the total hadronic cross section is calculated as a sum of
exclusive cross sections. However, in the energy region between 1.5-2.0 GeV 
the exclusive data are incomplete. In this section we discuss two previously
unstudied process.

\begin{figure}[ht]
\centering
\includegraphics[width=0.3\textwidth]{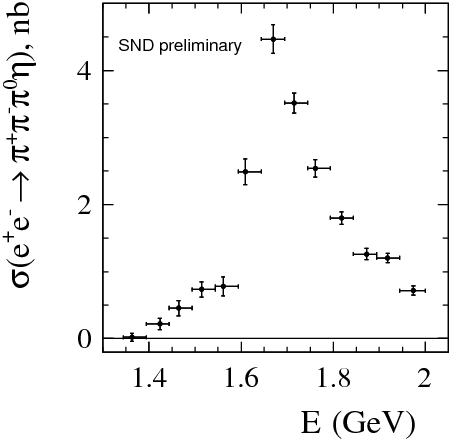}
\includegraphics[width=0.3\textwidth]{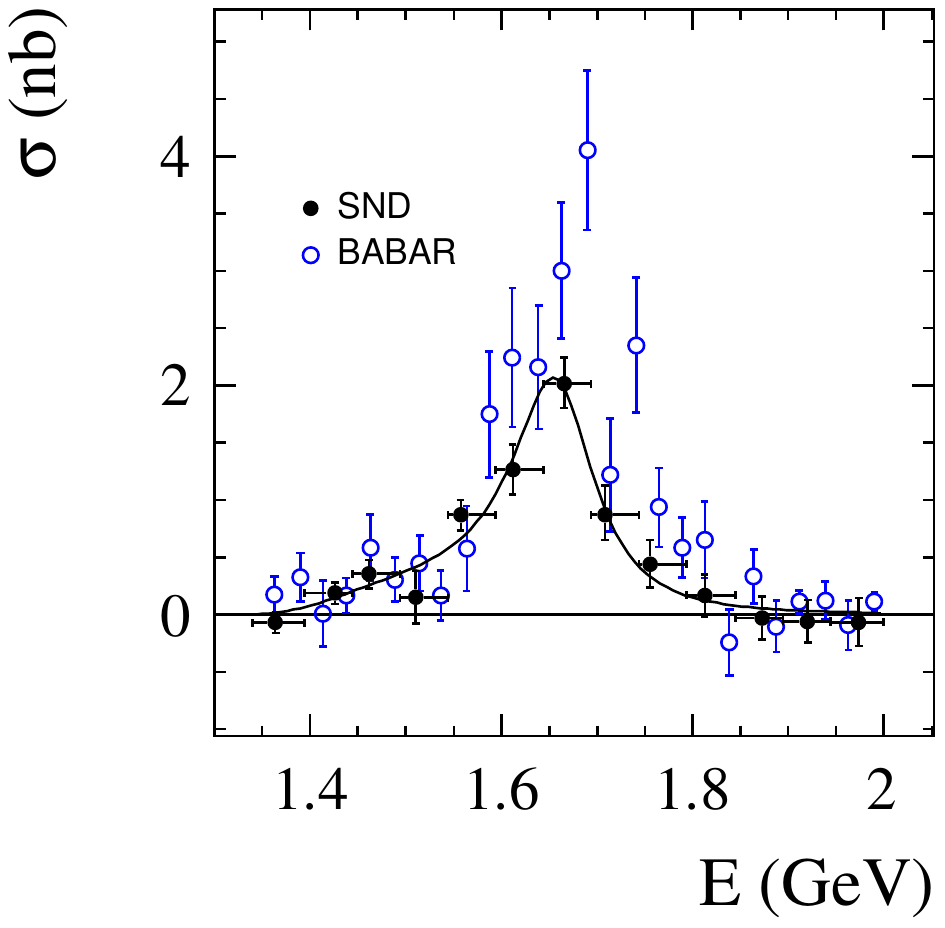}
\caption{Left panel: The $e^+e^-\to \pi^+\pi^-\pi^0\eta$ cross section measured
by SND.
Right panel: The $e^+e^- \to \omega \eta$ cross section measured by SND in
comparison with BABAR data~\cite{babar1}. The curve is the result of the VMD
fit.
\label{3piet}}
\end{figure}

The process $e^+e^-\to\pi^+\pi^-\pi^0\eta$ has complex internal structure.
Our preliminary study show that there are at least four mechanisms for 
this reaction: $\omega\eta$, $\phi\eta$, $a_0(980)\rho$, and 
structureless $\pi^+\pi^-\pi^0\eta$. The known $\omega\eta$ and $\phi\eta$ 
contributions explain about 50-60\% of the cross section below 1.8 GeV.
Above 1.8 GeV the dominant mechanism is $a_0\rho$.
The preliminary result on the $e^+e^-\to\pi^+\pi^-\pi^0\eta$ cross 
section is shown in Fig.~\ref{3piet} (left). The cross section
for the subprocess $e^+e^-\to \omega\eta$ is measured 
separately~\cite{ometa} and shown in Fig.~\ref{3piet} (right) in comparison
with the BABAR measurement~\cite{babar1}. Our results have better accuracy and
disagree with the BABAR data at $E >1.6$~GeV.

\begin{figure}[ht]
\centering
\includegraphics[width=0.3\textwidth]{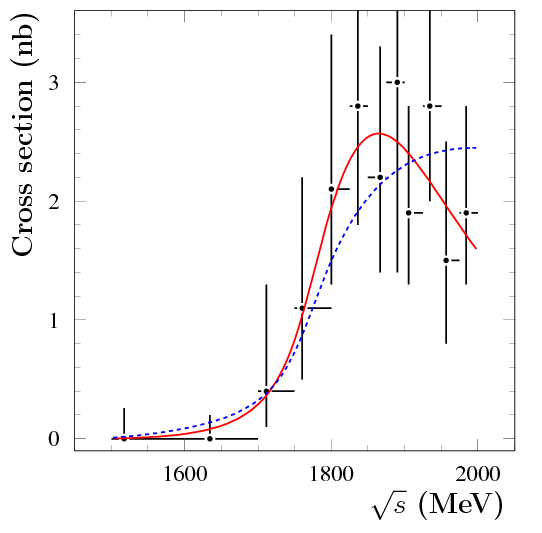}
\includegraphics[width=0.3\textwidth]{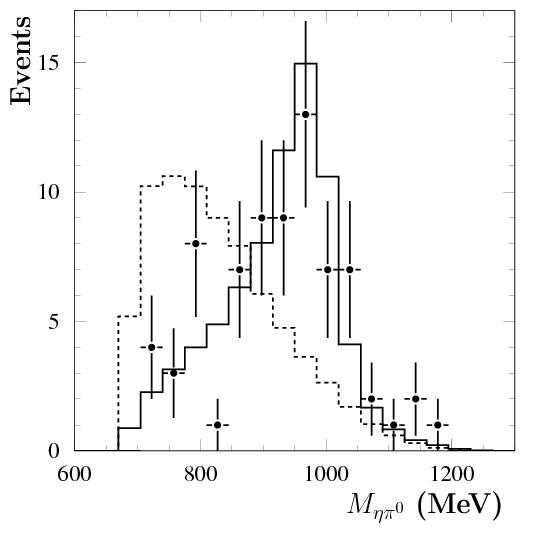}
\caption{Left panel: The $e^+e^-\to\omega\eta\pi^0$ cross section measured by
SND. The solid (dashed) curve shows the result of the fit in the model
of $\omega a_0(980)$ intermediate state with (without) a resonance contribution.
Right panel: The $\eta\pi^0$ invariant mass spectrum for selected 
$e^+e^-\to\omega\eta\pi^0$ events. The solid histogram represents
$e^+e^-\to\omega a_0(980)$ simulation, while the dashed histogram represents
$\omega\eta\pi^0$ phase-space simulation.
\label{csompiet}}
\end{figure}
The process $e^+e^-\to \omega\pi^0\eta$ is studied in the seven-photon
final state~\cite{etpig}. Events of the $e^+e^-\to \pi^0\pi^0\eta\gamma$ process
are selected. The analysis of their $\pi^0\gamma$ invariant-mass distribution
shows the dominance of the $\omega\pi^0\eta$ intermediate state. The measured
$e^+e^-\to \omega\pi^0\eta$ cross section is shown in 
Fig.~\ref{csompiet} (left). Figure~\ref{csompiet} (right) shows
the $\pi^0\eta$ mass distribution for selected $\omega\pi^0\eta$ events,
which is well described by the model the of $\omega a_0(980)$ intermediate 
state.

Both the previously unmeasured cross sections discussed in this section 
give a sizable contribution ($\sim5\%$) to the total
hadronic cross section.

\begin{acknowledgement}
This work is supported in part by the RFBR grants
16-02-00327, 16-02-00014, 15-02-01037, 16-32-00542, and 14-02-00129.
Part of this work related to the photon reconstruction algorithm in the
electromagnetic calorimeter is supported
by the Russian Science Foundation (project No. 14-50-00080).
\end{acknowledgement}
%
% BibTeX or Biber users please use (the style is already called in the class, ensure that the "woc.bst" style is in your local directory)
% \bibliography{name or your bibliography database}
%
% Non-BibTeX users please use
%

\end{document}